\documentclass[12pt]{article}
\usepackage{cite}
\newcommand{\be}{\begin{equation}}
\newcommand{\bea}{\begin{eqnarray}}
 
\newcommand{\eea}{\end{eqnarray}}
\newcommand{\ba}{\begin{array}}
\newcommand{\ea}{\end{array}}
\newcommand{\ee}{\end{equation}}

\expandafter\ifx\csname mathbbm\endcsname\relax

\else

\fi
\begin{document}
\begin{titlepage}
\hfill
\vbox{
    \halign{#\hfil         \cr
           IPM/P-2005/083 \cr
           } 
      }  
\vspace*{20mm}
\begin{center}
{\Large {\bf On Supergravity Solutions of Branes in Melvin Universes}\\ }

\vspace*{15mm}
\vspace*{1mm}
{Mohsen Alishahiha$^a$\footnote{alishah@theory.ipm.ac.ir}, 
Batool Safarzadeh$^b$\footnote{safarzadeh@theory.ipm.ac.ir}
and Hossein Yavartanoo$^a$\footnote{yavar@ipm.ir}
}
 \\
\vspace*{1cm}

{\it$^a$ Institute for Studies in Theoretical Physics 
and Mathematics (IPM)\\
P.O. Box 19395-5531, Tehran, Iran \\ \vspace{3mm}
$^b$  Department of physics, School of Science\\
Tarbiat Modares University, P.O. Box 14155-4838, Tehran, Iran}\\

\vspace*{1cm}
\end{center}

\begin{abstract}
We study supergravity solutions of type II branes wrapping a Melvin universe. These solutions
provide  the gravity description of non-commutative field theories with non-constant non-commutative
parameter. Typically these theories are non-supersymmetric, though they exhibit some feature
of their corresponding supersymmetric theories. 
An interesting feature of these non-commutative theories is that there is a critical length in the
theory in which for distances larger than this length the effects of non-commutativity become
important and for smaller distances these effects are negligible. Therefore we would expect to see
this kind of non-commutativity in large distances which might be relevant in cosmology.
We also study M5-brane wrapping on 11-dimensional Melvin universe and its descendant theories upon
compactifying on a circle.  

\end{abstract}

\end{titlepage}

\section{Introduction}
AdS/CFT correspondence \cite{{Maldacena:1997re},{Gubser:1998bc},{Witten:1998qj}}
have probably provided a powerful framework for 
understanding quantum gravity. In this framework a quantum mechanical system which includes gravity
can be described by a lower dimensional quantum mechanical system without gravity. Having had gravity on one side
of the duality one may then wonder if we can learn about quantum gravity by studying a quantum
field theory which by now we have more control on it. Of course, the point would be 
how to identify the two sides of the correspondence, namely starting from given  gravitional theory how to find 
the corresponding field theory dual. Without such an identification, it seems that although AdS/CFT
correspondence has opened up a window to understand the quantum gravity better,
practically it could not help us so much.

Fortunately string theory and different branes in string theory have been able to give us a practical way
to proceed and in fact by now we know several examples of AdS/CFT correspondence where we almost
know two sides of the duality. In general one could start from a given brane configuration
in string theory and check if the theory on the worldvolume of this brane configuration 
decouples from the bulk gravity in an special limit (decoupling limit). If so, one then
expect that string theory (gravity) on this particular background would be dual to the theory
lives on the worldvolume of the brane configuration.

The simplest example is to start from Dp-brane in type II string theories. It can be shown that
the brane worldvolume theory decouples from the bulk gravity for $p<6$ \cite{Alishahiha:2000qf}.
Therefore type II string theories in the near horizon limit (decoupling limit) of Dp-brane 
provide a gravity description for $(p+1)$-dimensional
gauge theory with 16 supercharges \cite{Itzhaki:1998dd}. In other words one may use these
SYM theories with 16 supercharges to study string theory/gravity on these particular backgrounds.
This procedure has also been generalized
for other branes like NS5-brane as well as M-theory branes. See for example \cite{Aharony:1998ub}.

Considering a single brane in string theory would probably mean that we are restricting ourselves in 
a small region of string theory moduli space. In generic point of string theory moduli space
we would expect different low energy fields have non-zero expectation value.
In particular in generic point we would expect to have non-zero NS-NS B-field.  
Turning on a B field on the D-brane worldvolume can be viewed,
via AdS/CFT correspondence, as a perturbation of the 
worldvolume field theory by an operator of dimension 6. 
For example in the D3-brane case, from the four dimensional 
superconformal Yang-Mills
theory point of view the bosonic part of this dimension 6 operator
is given by \cite{{Das:1998ei},{Ferrara:1998bp}}
\be
{\cal O}_{\mu\nu}={1\over 2g^2_{\rm YM}}{\rm Tr}\left(
F_{\mu \delta}F^{\delta \rho}F_{\rho \nu}
-F_{\mu\nu}F^{\rho\delta}F_{\rho \delta}+2
F_{\mu\rho}\sum_{i=1}^{6}\partial_{\nu}\phi^{i}
\partial^{\rho}\phi^{i}-{1\over 2}F_{\mu\nu}\sum_{i=1}^{6}
\partial_{\rho}\phi^{i}
\partial^{\rho}\phi^{i}\right)\;,
\ee
where $g_{\rm YM}$ is the SYM coupling, $F_{\mu\nu}$ is the 
$U(N)$ field strength and $\phi^i,\;i=1,\cdots, 6$ are the 
adjoint scalars. This deformed theory by
the operator ${\cal O}_{\mu \nu}$ can be extended to a complete 
theory with a simple description  which is non-commutative SYM theory.

In fact it has been shown in \cite{{Douglas:1997fm},{Ardalan:1998ce},{Chu:1998qz}}
 that, when we turn on a constant B field on the D-brane
worldvolume, the low-energy effective worldvolume theory is modified to be a
non-commutative Super-Yang-Mills (NCSYM) theory. Actually the worldvolume theory
of $N$ coincident Dp-branes in the presence of a B field is found
to be $U(N)$ NCSYM theory \cite{Seiberg:1999vs}.

As in the case of zero B-field, there exists a limit where the bulk
modes decouple from the worldvolume non-commutative field theory
\cite{Seiberg:1999vs}; we expect to have a correspondence between string theory
on the curved background with B field and non-commutative field theories.
In other words we expect to have a holographic picture like AdS/CFT 
correspondence  \cite{{Maldacena:1997re},{Gubser:1998bc},{Witten:1998qj}} 
for the non-commutative theories. In fact this issue has been investigated in several papers,
including \cite{{Hashimoto:1999ut},{Maldacena:1999mh},{Li:1999am},{Alishahiha:1999ci},{Harmark:1999rb},
{Barbon:1999mx},{Lu:1999rm},{Cai:2000hn},{Cai:1999aw},{Cai:2000yk}}.

So far we have consider the cases where the B-field is turned on in some spatial
directions along the brane worldvolume. One could also consider cases where B-field has one leg 
along the time direction. While space non-commutativity can be accommodated within field theory, 
space-time non-commutativity seems to require string theory for consistency
\cite{{Seiberg:2000ms},{Gopakumar:2000na},{Barbon:2000sg},{Gomis:2000zz},{Gopakumar:2000ep}}.
The B-field could also be light-like \cite{{Aharony:2000gz},{Alishahiha:2000pu},{Cai:2000py}}. One may also 
consider the worldvolume theory of a D-brane in the presence 
of non-zero B field with one leg along the brane worldvolume 
and the other along the transverse directions to the brane. 
This brane configuration  was studied in \cite{{Bergman:2000cw},{Chakravarty:2000qd},
{Dasgupta:2000ry},{Bergman:2001rw},{Dasgupta:2001zu},{Alishahiha:2002ex},{Alishahiha:2003ru}}
where the twisted
compactification was introduced. This twisted compactification 
leads us to introduce a new type of star product between
the fields at the level of effective field theory. The corresponding field theory is called 
dipole field theory.

One could also extend this consideration for NS5-brane/M-theory branes when 
we have non-zero RR field/3-form. In fact different deformations of 
NS5-brane with non-zero RR fields lead to theories 
on the  worldvolume of NS5 branes, whose 
excitations include light-open Dp branes (ODp) \cite{{Gopakumar:2000ep},{Alishahiha:2000pu},
{Harmark:2000ff},{Alishahiha:2000er}}.
 
So far we have considered those theories which can be arisen in the brane worldvolume when we have 
uniform B-field. One may also consider the cases where the B-field is not uniform. This could lead
to non-commutative field theories where the non-commutative parameter is non-constant  \cite{{Ho:2000fv},{Cornalba:2001sm},{Ho:2001fi},{Herbst:2001ai},{Cerchiai:2003yu},{Calmet:2003jv},
{Bertolami:2003nm},{Robbins:2003ry},{Behr:2003qc},{Das:2003kw},{Gayral:2005ih}}.
Several aspects of non-commutative field theory with non-constant (including time dependent B-field)
have been studied in \cite{{Dolan:2002px},{Hashimoto:2002nr},{Lowe:2003qy},{Hashimoto:2004pb},
{Alishahiha:2002bk},{Cai:2002sv},{Hashimoto:2005hy}}.
This is the aim of this paper to further study supergravity solution of type II string
theories in the presence of non-zero B-field which could provide  the gravity description
of non-commutative field theories with non-constant non-commutative parameter. We will
also extend this study for NS5-brane as well as M5-brane.

A common feature of adding B-field in the string theory (gravity) side of AdS/CFT correspondence is that
the corresponding field theory dual turns out to be a non-local theory. On the other hand if we are willing
to understand quantum gravity better, one would probably need to go beyond the local field theory. Therefore
studying of these non-local field theories could increase our knowledge about general properties of non-local field theories.

The organization of the paper is as follows. In section 2 we will review the
non-commutative gauge theory and the way we can write an invariant action when the
non-commutative parameter is non-constant. In section 3 we will obtain the supergravity
solution of Dp-brane wrapping a Melvin universe. 
In section 4 we shall study the decoupling limit of the supergravity 
solutions we have found in section 3. These could provide the gravity description of 
non-commutative gauge theory with a non-constant parameter in various dimensions.
In section 5 we will study type II NS5-brane wrapping a Melvin universe which leads to
new supergravity solutions of NS5-brane in the presence of different RR fields which depend on 
the brane worldvolume coordinates. In section 6 this procedure is generalized to M5-brane. 
By compactifying this solution  and bringing it to type IIA and then  using a chain of
T and S dualities we will generate new solutions representing Dp-brane in 
the presence of B-field with one leg along time direction. This is in fact the
generalization of NCOS theories where the non-commutative parameter is non-constant.
To complete our discussions we study the light-like deformation in section 7. The last 
section is devoted to discussions.

\section{Non-commutative gauge theory with\\ non-constant parameter}

In this section following \cite{Hashimoto:2005hy} 
we review non-commutative gauge theory with non-constant parameter.
We will consider a special case of non-commutativity which can be defined in the
worldvolume of D3-brane wrapping a Melvin universe. Although the non-commutative
parameter is not constant, one can still study the corresponding gauge theory using some
kind of star product. Of course it cannot be a simple Moyal product we usually use in the
non-commutative gauge theory when its parameter is constant. This is because, it cannot be
used to construct a gauge invariant action, taking into account that differentiation does
not respect the product rule with non-constant parameter.

Nevertheless it has been shown \cite{Hashimoto:2005hy} that in the polar 
coordinates the non-commutative parameter can be taken to be a constant. 
In fact for a four dimensional space parameterizing  by $t,r,\phi,x$ the non-commutative parameter 
could be taken constant which we denote it by 
$\theta^{\phi x}$. In this notation the star product is defined as
\be
f\# g=e^{\frac{i\theta^{\phi x}}{2}(\partial_\phi\partial_{x'}-\partial_{\phi'}\partial_{x})}
f(t,r,\phi,x)f(t,r,\phi',x')|_{\phi=\phi',x=x'}.
\ee
In this coordinates, one can define a set of unit vector fields as $\partial_i=X_a^\mu\partial_\mu$.
Explicitly we have
\be
\partial_1=\partial_t,\;\;\;\;\;\;\partial_2=\partial_r,\;\;\;\;\;\partial_3=
\frac{1}{r}\partial_\phi,\;\;\;\;\;\partial_4=\partial_x.
\ee
The $\#$ and $ * $ products are not related by a change of coordinates, though can be
related  using an automorphism  $R(f)$ \cite{Kontsevich:1997vb}
\be
R(f\# g)=R(f)*R(g),
\ee
where $R$ in leading order is given by \cite{Cerchiai:2003yu}
\be
R(f)=f+\frac{4\pi^2\theta^2}{24}r\partial_r\partial_x^2f+{\cal O}(\theta^3).
\ee
By making use of  this automorphism one can define a new derivation, which respects the product rule,
as follows
\be
\delta_{X_a}f=R\partial R^{-1}f.
\ee
Using this notation  it is now straightforward to write the action for the corresponding
non-commutative gauge theory \cite{Hashimoto:2005hy}
\be
S=\frac{1}{2}{\rm Tr}\int\sqrt{G}G^{ab}G^{cd}F_{ac}* F_{bd}, \;\;\;F_{ab}=\delta_{X_a}A_b-\delta_{X_b}
A_a+igA_a* A_b-igA_b* A_a,
\ee
where $G_{ab}=g_{\mu\nu}X^\mu_aX^\nu_b$. It can also be generalized for the case where we have 
scalar and spinor as well.

In the rest of this paper we shall study different aspects of non-commutative field theories defined by 
this non-commutative star product in various dimensions using their gravity duals.

\section{The Supergravity solution}

In this section we shall study the supergravity solution of Dp-brane wrapped in a Melvin universe in
type II string theories.\footnote{D-brane in Melvin universe has been studied in
\cite{{Takayanagi:2001aj},{Takayanagi:2001gu},{Dudas:2001ux}}.} 
This will lead, upon taking decoupling limit, to a non-commutative
gauge theory on the brane worldvolume with non-constant non-commutativity. To find this supergravity
solution we start from the supergravity solution of Dp-brane and performing a chain of T-dualities and
twists. In fact the procedure is very similar to one that studied \cite{Alishahiha:2003ru}
(see also \cite{Costa:1996zd}) in the
context of dipole field theory and we shall follow its notation.

The supergravity solution of $N$ coincident extremal Dp-branes in type II string theories 
in string frame is given by \cite{Horowitz:1991cd}
\bea
ds^2&=&f^{-1/2}(-dt^2+\sum_{i=1}^{p-1}dx_i^2+dx_p^2)+f^{1/2}(d\rho^2+\rho^2d\Omega_{8-p}^2),\cr &&\cr
e^{2\phi}&=&g_s^2f^{(3-p)/2},\;\;\;\;\;f=1+\frac{(2\pi)^{p-2}c_pNg_sl_s^{7-p}}{\rho^{7-p}},\;\;\;\;\;\;
C_{01\cdots p}=-\frac{1}{g_s}f^{-1},
\eea
where $c_p=2^{7-2p}\pi^{(9-3p)/2}\Gamma(\frac{7-p}{2})$. 

Suppose $x_p$ is compact with radius $R$. Setting $x_p=\beta_p\theta$ and performing a T-duality along  $\theta$ direction, one gets
\bea
ds^2&=&f^{-1/2}(-dt^2+\sum_{i=1}^{p-1}dx_i^2)+f^{1/2}(\frac{{\alpha'}^2}{\beta_p^2}d\tilde{\theta}^2+d\rho^2+\rho^2d\Omega_{8-p}^2),\cr &&\cr
e^{2\phi}&=&\frac{{\alpha'}^2}{\beta_p^2}g_s^2f^{(4-p)/2},\;\;\;\;\;\;\;C_{0\cdots (p-1)}=-f^{-1},
\eea
which corresponds to D$_{(p-1)}$-brane smeared along one direction, $\tilde{\theta}$, that is an
angular coordinate with period $\tilde{\theta}\sim\tilde{\theta}+2\pi$.  Now let us add a twist
to the direction along the brane worldvolume as we go around the circle $\tilde{\theta}$
\be
dx_i\rightarrow dx_i+\sum_j\Omega_{ij}x_jd\tilde{\theta},
\ee
where $\Omega_{ij}$ is an element of Lie algebra $so(p-1)$. Therefore the metric changes to
\be
ds^2=f^{-1/2}(-dt^2+\sum_{i=1}^{p-1}(dx_i+\Omega_{ij}x_jd\tilde{\theta})^2)+f^{1/2}(\frac{{\alpha'}^2}{\beta_p^2}d\tilde{\theta}^2+d\rho^2+\rho^2d\Omega_{8-p}^2).
\ee
It is useful to set a new notation in which $X^T=(x_1,\cdots,x_{p-1})$ and therefore the above metric reads
\bea
ds^2&=&f^{-1/2}(-dt^2+dX^TdX+2(X^T\Omega^TdX)d\tilde{\theta})
+(f^{-1/2}X^T\Omega^T\Omega X+f^{1/2}\frac{{\alpha'}^2}{\beta_p^2})
d\tilde{\theta}^2\cr &&\cr &+&f^{1/2}(d\rho^2+\rho^2d\Omega_{8-p}^2).
\eea
Finally, once again, we can apply another T-duality on $\tilde{\theta}$ direction. Doing so, in the limit 
of $\beta_p\rightarrow \infty$ while keeping $\beta_p\Omega=M$ and $x_p=\beta_p\theta$ fixed, one finds  
\bea
\label{sugra1}
ds^2&=&f^{-1/2}\left(-dt^2+dr^2+r^2dn^Tdn+\frac{{\alpha'}^2dx_p^2-r^4f^{-1}(n^TMdn)^2}{{\alpha'}^2+r^2f^{-1}n^TM^TMn}\right)\cr
 &+&f^{1/2}\left(d\rho^2+\rho^2d\Omega_{8-p}^2\right),\cr &&\cr
 e^{2\phi}&=&\frac{{\alpha'}^2g_s^2f^{(3-p)/2}}{{\alpha'}^2+r^2f^{-1}n^TM^TMn},\;\;\;\;\;
 \sum_i B_{pi}dx_i=\frac{r^2f^{-1}dn^TMn}{{\alpha'}^2+r^2f^{-1}n^TM^TMn}.
\eea
We will also get several RR fields which we have not written them here. 
We will write their explicit form when we consider each case in detail. 
Here we have also used a notation in which $X=rn$ for $n^Tn=1$.

It is worth noting that given a general supergravity solution of a system 
of branes, it is not clear whether the solution would give a well-defined 
description of some field theories. In fact, we must check to see 
whether there is a well-defined field theory on the brane worldvolume 
which decouples from bulk gravity. This can be done by evaluating the graviton absorption
cross section. If there is a limit where graviton absorption cross section vanishes, we have
a field theory which decouples from bulk gravity. Alternatively, one could calculate the 
potential that the gravitons feel because of the brane. Having a 
decoupled theory can be seen from the shape of the potential in 
the decoupling limit. Actually, for those branes which their 
worldvolume decouple from gravity, the potential develops an infinite 
barrier separating the space into two 
parts: bulk and brane. 
In this case the bulk's modes can not reach the brane because of this
infinite barrier, and the same for brane's modes. Therefore the theory on the brane 
decouples from the bulk. 

For the case we are interested in, perturbing the metric of the background (\ref{sugra1}), 
one finds the following equation for transverse gravitons \cite{Alishahiha:2000qf}
\be
\partial_{\mu} \left(\sqrt{-g} e^{- 2 \phi} g^{\mu\nu} \partial_{\nu}\Phi\right)=0,
\ee
with $\Phi=h(r) e^{i k_{\mu} x^{\mu}}$. From this equation we can read 
the potential by writing it in the form of a Schr\"odinger-like equation as follows
\be
\partial_{\rho}^2 \psi(\rho) +V_p(\rho) \psi(\rho)=0\;,
\ee 
where the potential is given in terms of the metric components and
in general it is messy to write the potential explicitly, though for the special
case where the twist acts just on two directions one can write it in a simple closed 
form which can give us an insight whether
the theory decouples. Doing so one arrives at  
\be V_p(\rho) = -\left(1+\frac{c_pNg_s(\omega l_s)^{7-p}}
{\rho^{7-p}}\right)+ \frac{(8-p)(6-p)}{4 \rho^2},
\ee 
with $\rho=\omega r$. This potential is the same as  the 
ordinary D-branes as well as branes in the presence of uniform B-field. 
Therefore we conclude that we have a decoupled  theory 
living on the worldvolume of Dp-brane for $p\leq 5$. Although we have not written
the potential for the most general twist, one can still show that in general the potential 
develops a barrier in the decoupling limit and thus we get decoupled theory.

\section{Supergravity description of non-commutative\\ gauge theory with non-constant parameter}

In the previous section we have shown that the worldvolume theory 
of Dp-brane in the presence of non-zero B-field given in (\ref{sugra1})
decouples from the gravity for $p\leq 5$. Therefore the solution (\ref{sugra1}) can 
provide a dual gravity description for non-commutative field theory with non-constant parameter
via AdS/CFT correspondence. 

For this supergravity solution the decoupling limit is defined as a limit in which
$\alpha'\rightarrow 0$ and keeping the following quantities 
fixed
\be
U = \frac{\rho}{l_s^2} , \;\;\;\;\;\;\;\;\;\; \bar{g_s} = g_s l_s^{p-3}.
\ee
In this limit the supergravity solution (\ref{sugra1}) reads
\bea
\label{sugra2}
l_s^{-2} ds^2&=& h^{1/2}\left(-dt^2+dr^2+r^2dn^Tdn+\frac{dx_p^2-r^4h(n^TMdn)^2}{1+r^2h (n^TM^TMn)}\right) \cr
 &+&h^{-1/2}\left(d\rho^2+\rho^2d\Omega_{8-p}^2\right),\cr &&\cr\cr
 e^{2\phi}&=&\frac{\bar{g}_s^2h^{(p-3)/2}}{1+r^2h(n^TM^TMn)},\;\;\;\;\; \sum_i B_{pi}dx_i=\frac{r^2h (n^TM^Tdn)}{1+r^2h(n^TM^TMn)},
\label{gen}
\eea
where 
\be
 R^{7-p} = 2^{7-2p}\pi^{(9-3p)/2}\Gamma(\frac{7-p}{2})g_{YM}^2 N ,\;\;\;\; g_{YM}^2 = (2\pi)^{p-2}\bar{g_s},
 \;\;\;\;h=\left(\frac{U}{R}\right)^{7-p}. 
\ee
The conjecture is now that the string theory on these backgrounds provides the
gravity description of non-commutative gauge theories with non-constant
non-commutative parameter in various dimensions. 

The effective dimensionless coupling constant in the corresponding 
non-commutative field theory can be defined as 
$g_{\rm eff}^2\sim g_{\rm YM}^2 N U^{p-3}$ and the scalar curvature of the metric in (\ref{gen}) has the 
behavior $l_s^2{\cal R}\sim \frac{1}{g_{\rm eff}}$. Thus the perturbative calculations in 
non-commutative field theory can be trusted when $g_{\rm eff}\ll 1$, 
while when  $g_{\rm eff}\gg 1$ the supergravity description is valid. 
We note also that the expression for dilaton in (\ref{gen}) can be recast to 
\be
e^{\phi}={1\over N}\;\frac{g_{\rm eff}^{(7-p)/2}}{(1+\frac{r^2(n^TM^TMn)U^{7-p}}{R^{7-p}})^{1/2}}\;.
\label{DIL}
\ee
Keeping $g_{\rm eff}$ and $r$ fixed we see from 
(\ref{DIL}) that $e^{\phi}\sim 1/N$. Therefore the string
loop expansion corresponds to $1/N$ expansion of non-commutative gauge 
theory.

Since the scalar curvature is $r$-independent, as far as the effective gauge coupling is concerned,
the situation is the same as ordinary brane solution. But since the dilaton is 
$r$-dependent this will change the phase structure
of the theory. In particular at given fixed energy the effective string 
coupling will change with $r$. There is, in fact, a critical length 
$r_c=g_{YM}\sqrt{N}/bU^{(7-p)/2}$ in which for $r\gg r_c$ the 
non-commutative effects become important  and the effective string
coupling becomes
\be
e^{2\phi}\sim \frac{ (g_{YM}^2N)^{(9-p)/2}}{N^2}\;\frac{U^{(p-5)(7-p)/2}}{b^2
r^2}\;,
\ee
where $b^2=n^TM^TMn$ is twist parameter. Therefore the effective string coupling decreases 
for large distance and the gravity description becomes more applicable.
On the other hand, at given fixed energy, the non-commutative effects become less
important for distances smaller than the critical length $r_c$.

One can also study the phase structure of the theory which is very
similar to that in non-commutative field theory with 
constant non-commutative parameter. The only difference is
that the distinguished points where the description of the theory
has to be changed is now $r$-dependent.

In the notation of \cite{Alishahiha:1999ci} the dimensionless effective 
non-commutative parameter is given by
\be
a^{\rm eff}= \left(\frac{r b U^2}{ g_{\rm eff}}\right)^{\frac{2}{ 7-p}}=\left(\frac{r}{r_c}\right)^{\frac{2}{7-p}}
\;.
\label{effnon}
\ee
At small distances
$r\ll r_c$ the non-commutative effects are small
and the effective description of the worldvolume theory is in terms
of a commutative field theory. Note that this distance is energy-dependent ($U$-dependent)
which means it changes with energy.

Form the expression of the  dimensionless effective non-commutative 
parameter (\ref{effnon}) one can read the non-commutative parameter  
seen by the gauge theory. In fact we get 
\be
[x_p,x_i]\sim {r b},
\ee
while in the polar coordinates it may be written as $[x_p,\theta]=b$ in agreement with 
\cite{Hashimoto:2005hy}.

To get a better insight of these
theories it is worth to study some of them in more detail. 

\subsection{D3-brane} 
This case has recently been studied in \cite{Hashimoto:2005hy}. Here, just for completeness, we will review this case
again. In our notation the corresponding matrix $M$ and unit vector $n$ are given by
\be
M=\pmatrix{0&b\cr -b&0},\;\;\;\;\;n^T=(\cos\theta\;\;\;\sin\theta).
\label{M3}
\ee
Plugging these into the general solution (\ref{sugra2}) we find
\bea
l_s^{-2} ds^2&=& \left(\frac{U}{R}\right)^2\left(-dt^2+dr^2 +
\frac{dx_3^2+r^2 d\theta^2}{1+\frac{r^2b^2U^4}{R^4}}\right)  + \left(\frac{R}{U}\right)^2
\left(dU^2+U^2d\Omega_{5}^2\right),\cr &&\cr\cr
 e^{2\phi}&=&\frac{\bar{g}_s^2}{1+\frac{r^2b^2U^4}{R^4}}, \;\;\;\;\;\;\;  B_{3\theta}=\alpha'
 \frac{r^2b\frac{U^4}{R^4}}{1+\frac{r^2b^2U^4}{R^4}},\;\;\;\;\;\;C_{0r}=\frac{\alpha'}{\bar{g}_s}br\frac{U^4}{R^4},
\label{d3}
\eea
where $R^4=2g_{YM}^2N$. We have also a RR 4-form corresponding to 
the original $N$ D3-branes which is given by
$dC_4=\frac{1}{\bar{g}_s}Nl_s^5\epsilon_5$ where $\epsilon_5$ is the worldvolume of the 5-sphere.

In spirit of AdS/CFT correspondence one may suspect that type IIB string theory in this background 
is dual to a non-commutative field theory with non-constant non-commutative parameter. 
The field
content of the theory is the same as ${\cal N}=4$ SYN theory in four dimension though 
the theory is not supersymmetric.

\subsection{D4-brane}
Let us consider D4-brane wrapping a Melvin universe. Using our general procedure the
corresponding solution is given by (\ref{gen}) with the following matrix $M$ and 
unit vector $n$
\be 
M=\pmatrix{0&0&0\cr 0&0&b \cr 0&-b&0},\;\;\;\;\;n^T=(\cos\theta\;\;\;\sin\theta\;\cos\phi\;\;\;\sin\theta\;
\sin\phi).
\label{M4}
\ee
Since the matrix $M$ is block diagonal one may work in the reduced subspace to simplify the computations.
To do that we consider a twist such that $x_1$ remains untouched, while
$x_2$ and $x_3$ transform the same as D3-brane case. Therefore we consider a twist which acts as follows
\be
(dx_1\;\;dx_2\;\;dx_3)=(dx_1\;\;dx_2+bx_3dx_4\;\;\;dx_3-bx_2dx_4).
\ee
In this notation we have $(x_2\;\;\;x_3)=r(\cos\theta\;\;\;\sin\theta)$ and the supergravity
solution reads
\bea
l_s^{-2} ds^2&=& \left(\frac{U}{R}\right)^{3/2}\left(-dt^2+dx_1^2+dr^2 +
\frac{dx_4^2+r^2 d\theta^2}{1+\frac{r^2b^2U^3}{R^3}}\right)  + \left(\frac{R}{U}\right)^{3/2}
\left(dU^2+U^2d\Omega_{4}^2\right),\cr &&\cr\cr
 e^{2\phi}&=&\bar{g}_s^2\frac{(U/R)^{3/2}}{1+\frac{r^2b^2U^3}{R^3}}, \;\;\;\;\;\;\; 
  B_{4\theta}=\alpha'\frac{r^2b\frac{U^3}{R^3}}{1+\frac{r^2b^2U^3}{R^3}},
  \;\;\;\;\;\;C_{01r}=\frac{{\alpha'}^{3/2}}{\bar{g}_s}br\frac{U^3}{R^3},
\label{d4}
\eea
where $R^3=g_{YM}^2N/4\pi$. There is also a RR 5-form representing the original $N$ D4-branes.

Actually this is the solution which could be obtained by using T-duality from D3-brane solution
(\ref{d3}). This solution could provide the gravity description of a gauge theory in five dimensions whose
field content is the same as five dimensional SYM theory with 16 supercharges, though the 
supersymmetry is broken because of non-zero B-field. One can then use this supergravity solution
to study the phase structure of the theory. In fact it can be seen that the phase structure is very
similar to the non-commutative field theory with constant non-commutative parameter. 
In particular the dilaton is small both in IR and UV limits and therefore in both limits
the type IIA supergravity solution provides a good description for the theory. 

\subsection{D5-brane }
In D5-brane case we recognize two different cases. The first case can simply be obtained from the D4-brane
solution (\ref{d4}) by a T-duality in a transverse direction to the brane. In this case the supergravity 
solution reads
\bea
l_s^{-2} ds^2&=& \frac{U}{R}\left(-dt^2+dx_1^2+dx_2^2+dr^2 +
\frac{dx_5^2+r^2 d\theta^2}{1+\frac{r^2b^2U^2}{R^2}}\right)  + \frac{R}{U}
\left(dU^2+U^2d\Omega_{3}^2\right),\cr &&\cr\cr
 e^{2\phi}&=&\bar{g}_s^2\frac{\frac{U^2}{R^2}}{1+\frac{r^2b^2U^2}{R^2}}, \;\;\;\;\;\;\;  B_{5\theta}=\alpha'
 \frac{r^2b\frac{U^2}{R^2}}{1+\frac{r^2b^2U^2}{R^2}},\;\;\;\;\;\;
 C_{012r}=\frac{{\alpha'}^2}{\bar{g}_s}br\frac{U^2}{R^2},
 \label{d51}
\eea
where $R^2=N\bar{g}_s$. We have also a RR 2-form corresponding to $N$ D5-branes.

On the other hand we can consider the most general case where all coordinates in the worldvolume of
the brane are touched by the twist. In this case the most general form for matrix $M$ up the a $so(4)$ 
transformation is given by 
\be 
M=\pmatrix{0&b&0&0\cr -b&0&0&0 \cr 0&0&0&b \cr 0&0&-b&0},
\label{M5}
\ee
and since the matrix is block diagonal the computations become simpler if we parametrize the 
unit vector $n$ as follows
\be
n^T=\pmatrix{\sin\theta\;\cos\phi &\cos\theta\;\cos\phi &\sin\psi\;\sin\phi &\cos\psi\;\sin\phi}.
\ee
With this parametrization the supergravity solution (\ref{gen}) reads
\bea
l_s^{-2} ds^2&=& \frac{U}{R}\left(-dt^2+dr^2 + r^2d\tilde{\Omega}_3^2 
+\frac{dx_5^2-r^4b^2\frac{U^2}{R^2}(\cos^2\phi d\theta+\sin^2\phi d\psi)^2}{1+\frac{r^2b^2U^2}{R^2}}\right) \cr &+& \frac{R}{U}\left(d\rho^2+\rho^2d\Omega_{3}\right),\cr &&\cr
e^{2\phi}&=&\bar{g}_s^2\frac{\frac{U^2}{R^2}}{1+\frac{r^2b^2U^2}{R^2}}, \;\;\;\;\;\;\;   
\sum_i B_{pi}d\theta_i=
\frac{r^2b\frac{U^2}{R^2}}{1+\frac{r^2b^2U^2}{R^2}}(\cos^2\phi d\theta+\sin^2\phi d\psi),
\eea
where $d\tilde{\Omega}_3^2=d\phi^2+\cos^2\phi d\theta^2+\sin^2\phi d\psi^2$. Beside the RR 2-form
representing the original $N$ D5-branes there is also a RR 4-form  which in the original $x_i$ coordinates 
is given by
\be
C_4=-b\frac{U^2}{R^2} dt\wedge\bigg{(}(x_1dx_1+x_2dx_2)\wedge dx_3\wedge dx_4+dx_1\wedge dx_2\wedge (x_3dx_3+x_4dx_4).
\bigg{)}
\ee

This supergravity solution provides a dual description of a six dimensional non-commutative gauge theory
with non-constant non-commutative parameter whose field content is the same as that in six dimensional 
SYM with 16 supercharges which is defined in the worldvolume of D5-brane, though because of non-constant 
B-field the supersymmetry is broken. One can then use this supergravity solution to study the corresponding
non-commutative gauge theory. In fact the phase digram of this system is very similar to
one with constant non-commutative parameter, though the typical scales where we will have
to change our description are $r$-dependent. At IR limit we expect that the non-commutative
effects become negligible, and therefore for $U\ll \sqrt{g_{YM}N}/rb$ the good description is given
by D5-brane solution without B-field and its phase digram would be the same as six dimensional 
supersymmetric gauge theory with 16 supercharges. On the other hand at UV limit where the 
effects of non-commutativity become important one needs to take into account the whole solution.
In particular in this limit the dilaton behaves like $e^{\phi}\sim \bar{g}_s/rb$ and we can trust
the gravity description as far as $\bar{g}_s\ll rb$, otherwise we need to make an S-duality and 
work in S-dual picture. In  this case we will have to deal with type IIB NS5-brane in the presence
of RR field which is the subject of the next section.

\section{NS5-brane wrapping a Melvin universe}

In this section we will study type II NS5-brane wrapping a Melvin universe. This
will lead to the supergravity solution of NS5-brane in the presence of several RR-fields.
This might be thought of as new deformation of the theories which live in the
worldvolume of NS5-brane in the presence of RR-field. When the deformation parameter 
is constant, these theories are known as ODp-theories  which include open Dp-branes.
On the other hand in the case we are interested in the deformation parameter is not constant 
and therefore we get new theories and this is the aim of this section to study these theories using
their dual supergravity solutions. 

These supergravity solutions can be obtained from D5-brane using a chain of S and T dualities.
As we saw in the previous section depending on how the twist acts on the coordinates, there are two 
different deformations of type IIB D5-brane. Starting from these solutions and apply S-duality 
one can find type IIB NS5-brane in the presence of RR 2-form.\footnote{ Under S-duality we have $\phi\rightarrow -\phi$ 
and $ds^2\rightarrow e^{-\phi}ds^2$ where $ds^2$ is the metric in string frame. Moreover the NSNS 
B-field gets changed to the RR 2-form.}  The simplest case is 
when the twist acts only on two coordinates (\ref{d51}).
Then a series of T-duality will generate other possible RR fields. Doing so one finds
\bea\label{NS5B}
ds^2&=& (1+r^2b^2f^{-1}/g_s^2\alpha'^2)^{1/2}\bigg{[}-dt^2+dr^2+\sum_{i=1}^{p-1}dx_i^2+
\frac{\sum_{a=p}^{4}dy_a^2}{1+r^2b^2f^{-1}/g_s^2\alpha'^2}  \cr
 &&\;\;\;\;\;\;\;\;\;\;\;\;\;\;\;\;\;\;\;\;\;\;\;\;\;\;\;\;\;\;\;\;\;\;\;\;+ f
(d\rho^2+\rho^2d\Omega_{3}^2)\bigg{]},\cr &&\cr
  C_{1\cdots(5-p)}&\sim&\frac{1}{{g}_s}\frac{r^2bf^{-1}}{1+r^2b^2f^{-1}/g_s^1\alpha'^2},
\;\;\;\;\;\;\;  
 C_{0r1\cdots (p-1)}\sim\frac{1}{{g}_s}brf^{-1},\cr &&\cr
e^{2\phi}&=&{g}_s^2(1+r^2b^2f^{-1}/g_s^2\alpha'^2)^{(p-1)/2}\frac{f}{r^{2\delta_{p5}}}.
\eea
Here in order to unify the solutions we have used a notation in which $dy_4=rd\theta,\;dx_4=d\theta/r$.
We have also a non-zero B-field representing the original  $N$ NS5-branes which in our notation is give by
$dB=Nl_s^2\epsilon_3$ with $\epsilon_3$ being volume of the 3-sphere.

We can also consider the decoupling limit of these solutions. The corresponding decoupling limit 
can be obtained from decoupling limit of D5-brane using S-duality in which $l_s^2\rightarrow g_sl_s^2$ and 
$g_s\rightarrow g_s^{-1}$. Thus the decoupling limit of the above supergravity solutions is defined as a 
limit in which $g_s\rightarrow 0$ while keeping the following quantities fixed
\be
U=\frac{\rho}{g_sl_s^2},\;\;\;\;\;\;\;l_s={\rm fixed}.
\ee
which is the same as the decoupling limit of little string theory \cite{Aharony:1998ub}. Note that
to make $U$ of dimension of energy, we have also added $l_s$ in the definition of $U$. In this
limit the supergravity solutions (\ref{NS5B}) read
\bea
ds^2&=& (1+\frac{r^2b^2U^2}{N\alpha'})^{1/2}\bigg{[}-dt^2+dr^2+\sum_{i=1}^{p-1}dx_i^2+
\frac{\sum_{a=p}^{4}dy_a^2}{1+\frac{r^2b^2U^2}{N\alpha'}} ,
 + \frac{N\alpha'}{\rho^2}
(d\rho^2+\rho^2d\Omega_{3}^2)\bigg{]},\cr &&\cr
e^{2\phi}&=&\frac{N}{\alpha' U^2}(1+\frac{r^2b^2U^2}{N\alpha'})^{(p-1)/2},\;\;\;\;\;\;
C_{0r1\cdots (p-1)}=\frac{{\alpha'}^{(p+1}/2}{{g}_s}\frac{brU^2}{N\alpha'},\cr &&\cr
&&C_{1\cdots(5-p)}=\frac{\alpha'^{(5-p)/2}}{{g}_s}\frac{\frac{r^2bU^2}{N\alpha'}}{1+\frac{r^2b^2U^2}{N\alpha'}}.
\eea
These supergravity solutions should be compared with supergravity solutions describing ODp-theories which are given
by NS5-brane in the presence of RR p-form. Here we also have the same structure though the RR fields are
also a function of the brane worldvolume coordinate $r$. This would result that the corresponding quantum theories
should be deformed by a non-constant parameter.

It is also constructive to study transverse gravitons scattering from these NS5-brane solutions in the 
presence of $r$-dependent RR fields. Ultimately this leads to a Schr\"ordinger-like equation
with the following potential
\be
V(\eta)=-1+(\frac{3}{4}-1)\frac{N\omega^2\alpha'}{\eta^2},
\ee
where $\eta=\omega\rho$ is a dimensionless radial coordinate. Following \cite{Minwalla:1999xi} one may
conclude that the theory has a mass gap of order of $m_{gap}\sim1/\sqrt{N\alpha'}$ which is exactly the same as 
six dimensional theories live on type II NS5-brane.

\section{M5-brane wrapping a Melvin universe and its descendant theories}

In this section we shall study supergravity solution of M5-brane wrapping 11-dimensional Melvin
universe. This would give a new non-commutative deformation of (0,2) theory with non-constant
non-commutative parameter. From gravity point of view this corresponds to the case where 
we have M5-brane in the presence of M-theory 3-form which depends on the coordinates of the
M5-brane worldvolume. In the case where the 3-form was independent of the brane worldvolume
coordinates, it was shown \cite{Gopakumar:2000ep} that one could consider a decoupling limit such 
that the theory in the M5-brane worldvolume decouples from bulk gravity and the decoupled theory
 has light open membrane. This theory is called OM theory. 

Upon compactifying OM theory on a circle and using 
a chain of T-duality, we will get supergravity solutions of Dp-brane in the
presence of electric E-field (B-field in the time direction). It was also shown \cite{{Seiberg:2000ms},{Gopakumar:2000na},{Barbon:2000sg},{Gomis:2000zz},{Gopakumar:2000ep}} 
unitarity implies that this theory is not a simple gauge theory and in fact the theory on the
corresponding worldvolume is indeed a non-commutative version of open string theory.
This is the aim of this section to generalize the above construction for the case where
the 3-form in M-theory and thereby the E-field in type II string theories depend on the
brane worldvolume coordinates. 

Let us first obtain the supergravity solution of M5-brane wrapping an 
eleven dimensional Melvin
universe. This can be done by making use of the type IIA supergravity
solution we have found.  Starting from D4-brane or type IIA NS5-brane one may uplift the solution to find
the M5-brane solution. In general a type IIA supergravity solution representing by 10-dimensional
metric, $ds_{10}^2$, RR one-form and dilaton can be uplifted into 11-dimensional solution
whose metric is given by
\bea 
ds_{11}^2= e^{4\phi/3}(dx_{11}+A_\mu dx^\mu)^2+e^{-2\phi/3}ds_{10}^2. 
\eea
Both RR 3-form and B-field under this uplifting go into M-theory 3-forms. 

Therefore to find the supergravity solution of M5-brane wrapping a Melvin universe, we can start
from D4-brane solution in (\ref{sugra1}) and then uplifting it to 11-dimensional supergravity and
sending the radius of 11$th$ direction to infinity $R_{11}\rightarrow \infty$. In this
limit keeping $bR_{11}=L$ fixed, one finds
\bea
ds^2&=& (1+r^2L^2f^{-1}/l_p^6)^{1/3}\bigg{[}f^{-1/3}\left(-dt^2+dr^2+dx^2 +
\frac{dy^2+dz^2+r^2 d\theta^2}{1+r^2L^2f^{-1}/l_p^6}\right)\cr &&\;\;\;\;\;\;\;\;\;\;\;\;\;\;\;\;\;\;
\;\;\;\;\;\;\;\;\;\;\;\;\;+ f^{2/3}
(d\rho^2+\rho^2d\Omega_{4}^2)\bigg{]},\cr &&\cr
 C_{yz\theta}&=&
 \frac{r^2Lf^{-1}}{1+r^2L^2f^{-1}/l_p^6},\;\;\;\;\;\;\;\;\;\;\;C_{0rx}= Lrf^{-1},\;\;\;\;\;\;\;\;\;\;
 f=1+\frac{\pi Nl_p^3}{\rho^3},
\eea
where $l_p$ is 11-dimensional Plank scale. There is also a 6-form (magnetic dual
to 3-form) representing the original $N$ M5-brane we started with. 
The decoupling limit of the solution is defined by $l_p\rightarrow 0$ while keeping $U=r/l_p^3$ fixed. 
One can easily write done the supergravity solution in this limit which will provide the 
gravity description of non-commutative (0,2) theory with non-constant non-commutative parameter.
One can also evaluate the curvature of the supergravity solution
\be
l_p^2{\cal R}\sim \frac{1}{N^{2/3}}\;\frac{1}{(1+\frac{r^2L^2U^3}{\pi N})^{1/3}},
\ee
which shows that we can trust the supergravity solution for large $N$. The supergravity can also
be trusted for scales much more larger that critical length defined in previous section.

One can easily check that upon compactifying this solution on one of a circle of $d\Omega_4^2$ 
we will end up with type IIA NS5-brane solution given in (\ref{NS5B}) for $p=2$. 
It is also possible to compactify it on other 
directions to find new solutions in type IIA string theory. One then use T-duality to find new 
solutions in type II string theories. Probably more interesting cases can by found by 
compactifying the M-theory solution on $x$ or $\theta$ and then using a chain S and T-dualities. 
For example compactifying on $x$ one finds
\bea
ds^2&=& (1+r^2b^2f^{-1}/\alpha'^2)^{1/2}\bigg{[}f^{-1/2}\left(-dt^2+dr^2 +
\frac{dy^2+dz^2+r^2 d\theta^2}{1+r^2b^2f^{-1}/\alpha'^2}\right)\cr &&\;\;\;\;\;\;\;\;\;\;\;\;\;\;\;\;\;\;
\;\;\;\;\;\;\;\;\;\;\;\;\;\;+ f^{1/2}
(d\rho^2+\rho^2d\Omega_{4}^2)\bigg{]},\cr &&\cr
 e^{2\phi}&=&g_s^2 f^{-1/2}(1+r^2b^2f^{-1}/\alpha'^2)^{1/2},\;\;\;\;\;B_{0r}= brf^{-1},\cr &&\cr
  C_{yz\theta}&=&\frac{1}{{g}_s} \frac{r^2bf^{-1}}{1+r^2b^2f^{-1}/\alpha'^2},\;\;\;\;\;\;\;\;\;
  \;\;\;\;\;
  f=1+\frac{\pi N g_sl_s^3}{\rho^3}.
\label{d4e}
\eea
This solution represent D4-brane supergravity solution in the presence of non-zero, non-constant
E-field ( B-field with one leg along the time direction). One can now proceed to find new solutions 
using T-duality. In fact T-dualizing the solution along $y$ we obtain D3-brane solution
smeared in one dimension. Then we can write down the localized D3-brane solution as follows
\bea
ds^2&=& (1+r^2b^2f^{-1}/\alpha'^2)^{1/2}\bigg{[}f^{-1/2}\left(-dt^2+dr^2 +
\frac{dz^2+r^2 d\theta^2}{1+r^2b^2f^{-1}/\alpha'^2}\right)\cr &&\;\;\;\;\;\;\;\;\;\;\;\;\;\;\;\;\;\;
\;\;\;\;\;\;\;\;\;\;\;\;\;\;+ f^{1/2}
(d\rho^2+\rho^2d\Omega_{5}^2)\bigg{]},\cr &&\cr
 e^{2\phi}&=&g_s^2 (1+r^2b^2f^{-1}/\alpha'^2),\;\;\;\;\;B_{0r}= brf^{-1},\cr &&\cr
  C_{z\theta}&=&\frac{1}{{g}_s} \frac{r^2bf^{-1}}{1+r^2b^2f^{-1}/\alpha'^2},\;\;\;\;\;\;\;\;\;
  \;\;\;\;\;
  f=1+\frac{4\pi N g_sl_s^4}{\rho^4}.
\eea
Doing the same in $z$ direction we arrive at the following D2-brane solution\footnote{
This solution can also be uplifted into M-theory to get M2-brane in the presence
of non-constant 3-form with two legs along the brane worldvolume and one leg
along the transever direction to the brane. This could be used to study a new deformation of
3-dimension ${\cal N}=8$ SCFT.}  
\bea
ds^2&=& (1+r^2b^2f^{-1}/\alpha'^2)^{1/2}\bigg{[}f^{-1/2}\left(-dt^2+dr^2 +
\frac{r^2 d\theta^2}{1+r^2b^2f^{-1}/\alpha'^2}\right)\cr &&\;\;\;\;\;\;\;\;\;\;\;\;\;\;\;\;\;\;
\;\;\;\;\;\;\;\;\;\;\;\;\;\;+ f^{1/2}
(d\rho^2+\rho^2d\Omega_{6}^2)\bigg{]},\cr &&\cr
 e^{2\phi}&=&g_s^2 f^{1/2} (1+r^2b^2f^{-1}/\alpha'^2)^{3/2},\;\;\;\;\;B_{0r}= brf^{-1},\cr &&\cr
  C_{\theta}&=&\frac{1}{{g}_s} \frac{r^2bf^{-1}}{1+r^2b^2f^{-1}/\alpha'^2},\;\;\;\;\;\;\;\;\;
  \;\;\;\;\;
  f=1+\frac{c_2 N g_sl_s^5}{\rho^5}.
\eea
Finally one could perform a T-duality along $\theta$ direction to find D1-brane solution as follows
\bea
ds^2&=& (1+r^2b^2f^{-1}/\alpha'^2)^{1/2}\bigg{[}f^{-1/2}(-dt^2+dr^2 )+ f^{1/2}
(d\rho^2+\rho^2d\Omega_{7}^2)\bigg{]},\cr &&\cr
 e^{2\phi}&=&g_s^2\frac{f}{r^2} (1+r^2b^2f^{-1}/\alpha'^2)^{2},\;\;\;\;\;B_{0r}= brf^{-1},\cr &&\cr
  \chi&=&\frac{1}{{g}_s} \frac{r^2bf^{-1}}{1+r^2b^2f^{-1}/\alpha'^2},\;\;\;\;\;\;\;\;\;
  \;\;\;\;\;
  f=1+\frac{c_1 N g_sl_s^6}{\rho^6}.
\eea
where $\chi$ is type IIB RR scaler. In all solutions we have an extra 
RR $p$-field representing
the original $N$ Dp-branes.

It is also possible to perform a T-duality along a direction transverse to the brane in the solution
(\ref{d4e}). Doing so we find a D5-brane solution in the presence of E-field which could also been
obtained from solution (\ref{NS5B}) for $p=1$ using S-duality.

These solutions, upon taking decoupling limit, must be compared with non-commutative open string theory
\cite{{Seiberg:2000ms},{Gopakumar:2000na},{Barbon:2000sg},{Gomis:2000zz},{Gopakumar:2000ep}}. We note,
however, that the decoupling limit of these solutions is the same as the case when we have B-field,
namely $l_s\rightarrow 0$ while $g_sl_s^{p-3}$ fixed. 

One the other hand in the case of constant non-commutative parameter, having E-field would cause
to have non-commutativity in the time direction and theory would be ill-defined unless 
we add open string in the game. This was automatically the case by taking near
critical E-field. But in our case, at least as far as the supergravity is concerned, the
decoupling limit is the same as the one with B-field.

If in this case the theory is going to be a non-commutative theory with
non-commutative time, one might suspect that the dual theory is not unitary, unless we could add
open string in the game. Otherwise, the theory would be ill-defined and from gravity point of view
this could mean that  the gravity solutions are unstable.
It would be interesting to study this case in more detail.

Finally we note that if we compactify the solution on $\theta$ direction one finds
\bea
ds^2=r\bigg{[}f^{-1/2}\left(-dt^2+dr^2+dx^2+\frac{dy^2+dz^2}{1+r^2b^2f^{-1}/\alpha'^2}\right)+f^{1/2}
(d\rho^2+\rho^2d\Omega_4^2)\bigg{]},\cr &&\cr
e^{2\phi}=g_s^2\frac{r^3f^{-1/2}}{1+r^2b^2f^{-1}/\alpha'^2}\;\;\;
B_{yz}=\frac{r^2bf^{-1}}{1+r^2b^2f^{-1}/\alpha'^2},\;\;\;C_{0rx}=\frac{1}{g_s}brf^{-1}.
\eea
Performing a T-duality along $x$ direction we will get another D3-brane solution in the 
presence of non-zero non-uniform B-field as follows
\bea
ds^2=r\bigg{[}f^{-1/2}\left(-dt^2+dr^2+\frac{dy^2+dz^2}{1+r^2b^2f^{-1}/\alpha'^2}\right)+f^{1/2}
(d\rho^2+\rho^2d\Omega_4^2)\bigg{]},\cr &&\cr
e^{2\phi}=g_s^2\frac{r^2}{1+r^2b^2f^{-1}/\alpha'^2}\;\;\;
B_{yz}=\frac{r^2bf^{-1}}{1+r^2b^2f^{-1}/\alpha'^2},\;\;\;C_{0r}=\frac{1}{g_s}brf^{-1}.
\eea
This solution upon taking the decoupling limit will provide the gravity description for a
non-commutative gauge theory with non-constant non-commutative parameter which should not be the
same as what we have reviewed in section 2. 

\section{Light-like twist}

In this section, for completeness, we will study Dp-brane wrapping a background with 
light-like twist.  The supergravity solution of light-like twist has been considered
in \cite{Alishahiha:2003ru}. In this section we generalize this construction for the case
where the light-like B-field depends on the brane worldvolume coordinates.

To find the corresponding solution we start from the general solution
(\ref{sugra1}) and consider the following boost in the $x_p$ direction
\be {\hat
t}=\cosh\gamma\;t-\sinh\gamma\;x_p,\;\;\;\;\; {\hat
x}_p=-\sinh\gamma\;t-\cosh\gamma\;x_p\;, \label{BBOOST1} 
\ee or
\be x^+=e^{-\gamma}y^+,\;\;\;\;\;x^{-}=e^{\gamma}y^-\;,
\label{BBOOST2} 
\ee 
with $y^{\pm}=x_p\pm t$ and $x^{\pm}={\hat
x}_p\pm {\hat t}$.

To have a light-like limit we now take the infinite boost limit,
$\gamma \rightarrow \infty$. In order to end up with a light-like
limit vector with finite component we must simultaneously scale
$M\rightarrow 0$ while $M e^{\gamma}=\tilde{M}$ is kept fixed. In this limit the
background (\ref{sugra1}) reads
\bea ds^2&=&f^{-1/2}\left(-4dx^+dx^--\frac{r^2f^{-1}}{\alpha'^2}(n^T\tilde{M}^T\tilde{M}n)
(dx^+)^2+dr^2+
r^2dn^T dn \right)\cr &&\cr &+&f^{1/2}\left( d
U^2+ U^2 d\Omega_{8-p}^2 \right),\cr&&\cr
e^{2\phi}&=&g_s^2f^{(3-p)/2},\;\;\;\;\;\;\;\ \sum_i B_{+i}dx_i=r^2f^{-1}(dn^T\tilde{M}n).
\eea
We have also an extra RR p-form representing $N$ Dp-branes. 

Similarly we can also apply this procedure for other solutions we have found in this paper.
Upon taking the decoupling limit these solution would provide the supergravity description
for different theories, in various dimensions with light-like non-commutative deformation with
non-constant non-commutative parameter. 

\section{Discussions} 

In this paper we have obtained supergravity solutions of different branes in type II string theories and M-theory
wrapping a Melvin universe. Practically these solutions can be obtained by a chain of T and S dualities and twists. 
These supergravity solutions correspond to Dp-brane in presence of non-zero B-field along its worldvolume 
such that the B-field depends on the brane worldvolume coordinates (non-constant). 
Doing the same procedure for NS5-brane
we have found a class of supergravity solutions corresponding to type II NS5-branes in the presence of different
RR fields along the brane worldvolume which are coordinates dependent. 

The supergravity solution of M5-brane  in the presence of non-zero, non-constant 3-form along
the worldvolume of the brane has also been obtained. Upon compactifying this solution on a circle,
depending on which direction is taken to be compact, and also using a chain of T and S dualities
we have been able to find new supergravity solutions corresponding to different brane solutions in 
presence of B-field with one leg along time direction. We have also considered a light-like
B-field/RR field which can be obtained from the solutions we have studied in the previous sections
by making use of an infinite boost.

We have seen that there is a limit in which the worldvolume theory of these solutions decouples
from the bulk gravity and therefore they could provide supergravity description of new deformation 
of the brane worldvolume theory.  In fact the situation is very similar to the case when 
we have Dp/NS5/M5 branes in the presence of non-zero B-field/RR field/3-form
which was independent of the brane worldvolume coordinates (constant). The worldvolume theory decouples 
from the bulk and therefore would provide a gravity description of non-commutative gauge 
theory/ODp-theory/OM-theory. 
It is also known that upon compcatifying OM-theory on a circle and using T and S dualities one will
get NCOS theory. From supergravity point of view this corresponds to Dp-brane in the presence of
E-field.

Therefore one may conclude that the supergravity solution we have found in this paper would provide
the gravity description of non-commutative deformation of the corresponding theories where the
non-commutative parameter is non-constant. 

In general we would expect that by turning on a non-zero, non-constant B-field in the worldvolume
on Dp-brane, the worldvolume  theory deforms to a non-commutative gauge theory with non-constant 
non-commutativity parameter. Because of non-constant B-field these theories are not supersymmetric,
nevertheless the field content of them are the same as their undeform supersymmetric theories.
In IR limit where the non-commutative effects are negligible, the supersymmetry restores. Since
the theory is not supersymmetric, one may wonder if the corresponding supergravity solution
is stable. This is a point one needs to be check, though we have not studied it in this paper.

One interesting feature of non-commutative field theories with non-constant parameter is that
they have a nature critical length which controls the effects of non-commutativity. As we have
seen, from supergravity description point of view, the non-commutative effects are controlled by a
dimensionless parameter given by
\be
a^{\rm eff}= \left(\frac{r b U^2}{ g_{\rm eff}}\right)^{\frac{2}{ 7-p}}=
\left(\frac{r}{r_c}\right)^{\frac{2}{7-p}}.
\ee
Therefore the non-commutative effects are important at  distances which are of order of $r_c$ while
they are negligible for distances smaller than this natural length. This is an interesting fact, 
saying that, we would expect to see non-commutative effects at large scaler in contrast to
our standard intuition and what we have learned in the case of non-commutative field theory
with constant parameter where we expect to see the effects at short distances.

To be concrete let us consider the D3-brane case in more detail. In fact an interesting feature about 
the critical length, $r_c$, is that it is a function of $U$, namely
\be
r_c=\frac{\sqrt{g_sN}}{bU^2}.
\label{qwe}
\ee
According to AdS/CFT correspondence one may think about coordinate $U$ as the
scale of energy and therefore the critical length is energy dependent parameter. 
At any given fixed energy, the non-commutative effects are given in terms of critical 
length (\ref{qwe}). On the other hand if we consider an $s$-wave scaler field $\Phi$ with frequency
$\omega$ in the background (\ref{d3}) the wave equation is
\be
U^{-3}\partial_U(U^{-5}\partial_U\Phi)+\omega^2\frac{g_{YM}^2N}{U^2}\Phi=0,
\ee
which shows that the solution only depends on $\omega^2\frac{g_{YM}^2N}{U^2}$ and so 
the radial dependence of the solution has the holographic relation with energy. Actually this means 
that a UV cutoff $U$ on radius of $AdS_5$ translates into a UV cutoff ${\cal E}$ in the dual CFT, 
such that
\cite{{Susskind:1998dq},{Peet:1998wn}}
\be
{\cal E}=\frac{U}{\sqrt{g_sN}}.
\ee 
Alternatively one would say that at the energy scale $U$ in the bulk, only those modes in CFT
will be excited which are in region given by $\delta X=\frac{\sqrt{g_sN}}{U}$
which is very similar to the relation we get for critical length. In fact these two can be combined
to get a limit on the non-commutativity
\be
r_c=\frac{1}{\sqrt{g_sN}}\;\frac{(\delta X)^2}{b},
\ee
where $\delta X$ would be a typical length of our normal life which could be of order of meter.
On the other hand the non-commutativity effects are important in the distances of order of $r_c$ and 
since we have not seen these effects so far, therefore at least $r_c$ must be of order of a typical 
cosmological length, or the radius of the world which is of order of $10^{26}\; m$. 
Putting this information as an input one may
put a bound on $b$ or to be precise on $\sqrt{g_s}b\sim 10^{-27}\;m$ if we assume $N$ is of order of
$10^2$.

To summarize we note that the non-commutative effects with non-constant parameter could affect
the long distance physics and therefore might be relevant in cosmology. It would be interesting
to study a cosmological model with such a non-commutativity and would probably put a bound 
on the non-commutative parameter using WMAP data.

To understand the feature of these kind of non-commutative field theory one could also study other
object in this theory like Wilson loop, monopoles and other salitonic solutions using AdS/CFT
correspondence dictionary. In particular one can use the open string action in this background to study
Wilson loop and thereby the effective potential between the external objects like ``quarks''
following \cite{{Maldacena:1998im},{Rey:1998ik}}. In fact in this case the situation is very similar to 
the case where the non-commutative parameter was constant and actually we get the same expression
as what studied in \cite{Alishahiha:1999ci} except that now the final results are $r$-dependent.

\noindent\textbf{Acknowledgments}

We would like to thank Farhad Ardalan for useful discussions.


\begin{thebibliography}{99}


\bibitem{Maldacena:1997re}
  J.~M.~Maldacena,
   ``The large N limit of superconformal field theories and supergravity,''
 Adv.\ Theor.\ Math.\ Phys.\  {\bf 2}, 231 (1998)
  [Int.\ J.\ Theor.\ Phys.\  {\bf 38}, 1113 (1999)]
  [arXiv:hep-th/9711200].


\bibitem{Gubser:1998bc}
  S.~S.~Gubser, I.~R.~Klebanov and A.~M.~Polyakov,
   ``Gauge theory correlators from non-critical string theory,''
    Phys.\ Lett.\ B {\bf 428}, 105 (1998)
  [arXiv:hep-th/9802109].

\bibitem{Witten:1998qj}
  E.~Witten,
   ``Anti-de Sitter space and holography,''
  Adv.\ Theor.\ Math.\ Phys.\  {\bf 2}, 253 (1998)
  [arXiv:hep-th/9802150].


\bibitem{Alishahiha:2000qf}
  M.~Alishahiha, H.~Ita and Y.~Oz,
  ``Graviton scattering on D6 branes with B fields,''
  JHEP {\bf 0006}, 002 (2000)
  [arXiv:hep-th/0004011].

\bibitem{Itzhaki:1998dd}
  N.~Itzhaki, J.~M.~Maldacena, J.~Sonnenschein and S.~Yankielowicz,
   ``Supergravity and the large N limit of theories with sixteen
   supercharges,''
  Phys.\ Rev.\ D {\bf 58}, 046004 (1998)
  [arXiv:hep-th/9802042].


\bibitem{Aharony:1998ub}
  O.~Aharony, M.~Berkooz, D.~Kutasov and N.~Seiberg,
   ``Linear dilatons, NS5-branes and holography,''
  JHEP {\bf 9810}, 004 (1998)
  [arXiv:hep-th/9808149].

\bibitem{Das:1998ei}
  S.~R.~Das and S.~P.~Trivedi,
   ``Three brane action and the correspondence between N = 4 Yang Mills  theory
   and anti de Sitter space,''
    Phys.\ Lett.\ B {\bf 445}, 142 (1998)
  [arXiv:hep-th/9804149].

\bibitem{Ferrara:1998bp}
  S.~Ferrara, M.~A.~Lledo and A.~Zaffaroni,
   ``Born-Infeld corrections to D3 brane action in AdS(5) x S(5) and N = 4,  d =
   4 primary superfields,''
    Phys.\ Rev.\ D {\bf 58}, 105029 (1998)
  [arXiv:hep-th/9805082].


\bibitem{Douglas:1997fm}
  M.~R.~Douglas and C.~M.~Hull,
   ``D-branes and the noncommutative torus,''
    JHEP {\bf 9802}, 008 (1998)
  [arXiv:hep-th/9711165].


\bibitem{Ardalan:1998ce}
  F.~Ardalan, H.~Arfaei and M.~M.~Sheikh-Jabbari,
  ``Noncommutative geometry from strings and branes,''
  JHEP {\bf 9902}, 016 (1999)
  [arXiv:hep-th/9810072].

\bibitem{Chu:1998qz}
  C.~S.~Chu and P.~M.~Ho,
  ``Noncommutative open string and D-brane,''
  Nucl.\ Phys.\ B {\bf 550}, 151 (1999)
  [arXiv:hep-th/9812219].





\bibitem{Seiberg:1999vs}
  N.~Seiberg and E.~Witten,
   ``String theory and noncommutative geometry,''
  JHEP {\bf 9909}, 032 (1999)
  [arXiv:hep-th/9908142].

\bibitem{Hashimoto:1999ut}
  A.~Hashimoto and N.~Itzhaki,
   ``Non-commutative Yang-Mills and the AdS/CFT correspondence,''
  Phys.\ Lett.\ B {\bf 465}, 142 (1999)
  [arXiv:hep-th/9907166].

\bibitem{Maldacena:1999mh}
  J.~M.~Maldacena and J.~G.~Russo,
   ``Large N limit of non-commutative gauge theories,''
  JHEP {\bf 9909}, 025 (1999)
  [arXiv:hep-th/9908134].

\bibitem{Li:1999am}
  M.~Li and Y.~S.~Wu,
   ``Holography and noncommutative Yang-Mills,''
  Phys.\ Rev.\ Lett.\  {\bf 84}, 2084 (2000)
  [arXiv:hep-th/9909085].

\bibitem{Alishahiha:1999ci}
  M.~Alishahiha, Y.~Oz and M.~M.~Sheikh-Jabbari,
  ``Supergravity and large N noncommutative field theories,''
  JHEP {\bf 9911}, 007 (1999)
  [arXiv:hep-th/9909215].

\bibitem{Harmark:1999rb}
  T.~Harmark and N.~A.~Obers,
   ``Phase structure of non-commutative field theories and spinning brane  bound
   states,''
    JHEP {\bf 0003}, 024 (2000)
  [arXiv:hep-th/9911169].

\bibitem{Barbon:1999mx}
  J.~L.~F.~Barbon and E.~Rabinovici,
   ``On $1/N$ corrections to the entropy of noncommutative Yang-Mills  theories,''
    JHEP {\bf 9912}, 017 (1999)
  [arXiv:hep-th/9910019].

\bibitem{Lu:1999rm}
  J.~X.~Lu and S.~Roy,
   ``$(p+1)$-dimensional noncommutative Yang-Mills and D$(p-2)$ branes,''
    Nucl.\ Phys.\ B {\bf 579}, 229 (2000)
  [arXiv:hep-th/9912165].

\bibitem{Cai:2000hn}
  R.~G.~Cai and N.~Ohta,
   ``Noncommutative and ordinary super Yang-Mills on (D$(p-2)$,D$p$) bound
   states,''
   JHEP {\bf 0003}, 009 (2000)
  [arXiv:hep-th/0001213].


\bibitem{Cai:1999aw}
  R.~G.~Cai and N.~Ohta,
  ``On the thermodynamics of large N non-commutative super Yang-Mills
  theory,''
  Phys.\ Rev.\ D {\bf 61}, 124012 (2000)
  [arXiv:hep-th/9910092].

\bibitem{Cai:2000yk}
  R.~G.~Cai and N.~Ohta,
  ``(F1, D1, D3) bound state, its scaling limits and SL(2,Z) duality,''
  Prog.\ Theor.\ Phys.\  {\bf 104}, 1073 (2000)
  [arXiv:hep-th/0007106].




\bibitem{Seiberg:2000ms}
  N.~Seiberg, L.~Susskind and N.~Toumbas,
   ``Strings in background electric field, space/time noncommutativity  and a
   new noncritical string theory,''
   JHEP {\bf 0006}, 021 (2000)
  [arXiv:hep-th/0005040].

\bibitem{Gopakumar:2000na}
  R.~Gopakumar, J.~M.~Maldacena, S.~Minwalla and A.~Strominger,
   ``S-duality and noncommutative gauge theory,''
   JHEP {\bf 0006}, 036 (2000)
  [arXiv:hep-th/0005048].

\bibitem{Barbon:2000sg}
  J.~L.~F.~Barbon and E.~Rabinovici,
   ``Stringy fuzziness as the custodian of time-space noncommutativity,''
   Phys.\ Lett.\ B {\bf 486}, 202 (2000)
  [arXiv:hep-th/0005073].

\bibitem{Gomis:2000zz}
  J.~Gomis and T.~Mehen,
   ``Space-time noncommutative field theories and unitarity,''
   Nucl.\ Phys.\ B {\bf 591}, 265 (2000)
  [arXiv:hep-th/0005129].

\bibitem{Gopakumar:2000ep}
  R.~Gopakumar, S.~Minwalla, N.~Seiberg and A.~Strominger,
   ``OM theory in diverse dimensions,''
   JHEP {\bf 0008}, 008 (2000)
  [arXiv:hep-th/0006062].

\bibitem{Aharony:2000gz}
  O.~Aharony, J.~Gomis and T.~Mehen,
   ``On theories with light-like noncommutativity,''
    JHEP {\bf 0009}, 023 (2000)
  [arXiv:hep-th/0006236].


\bibitem{Alishahiha:2000pu}
  M.~Alishahiha, Y.~Oz and J.~G.~Russo,
  ``Supergravity and light-like non-commutativity,''
  JHEP {\bf 0009}, 002 (2000)
  [arXiv:hep-th/0007215].

\bibitem{Cai:2000py}
  R.~G.~Cai and N.~Ohta,
  ``Lorentz transformation and light-like noncommutative SYM,''
  JHEP {\bf 0010}, 036 (2000)
  [arXiv:hep-th/0008119].

\bibitem{Bergman:2000cw}
  A.~Bergman and O.~J.~Ganor,
   ``Dipoles, twists and noncommutative gauge theory,''
   JHEP {\bf 0010}, 018 (2000)
  [arXiv:hep-th/0008030].

\bibitem{Chakravarty:2000qd}
  S.~Chakravarty, K.~Dasgupta, O.~J.~Ganor and G.~Rajesh,
   ``Pinned branes and new non Lorentz invariant theories,''
   Nucl.\ Phys.\ B {\bf 587}, 228 (2000)
  [arXiv:hep-th/0002175].

\bibitem{Dasgupta:2000ry}
  K.~Dasgupta, O.~J.~Ganor and G.~Rajesh,
   ``Vector deformations of N = 4 super-Yang-Mills theory, pinned branes,  and
   arched strings,''
   JHEP {\bf 0104}, 034 (2001)
  [arXiv:hep-th/0010072].

\bibitem{Bergman:2001rw}
  A.~Bergman, K.~Dasgupta, O.~J.~Ganor, J.~L.~Karczmarek and G.~Rajesh,
   ``Nonlocal field theories and their gravity duals,''
   Phys.\ Rev.\ D {\bf 65}, 066005 (2002)
  [arXiv:hep-th/0103090].

\bibitem{Dasgupta:2001zu}
  K.~Dasgupta and M.~M.~Sheikh-Jabbari,
   ``Noncommutative dipole field theories,''
    JHEP {\bf 0202}, 002 (2002)
  [arXiv:hep-th/0112064].

\bibitem{Alishahiha:2002ex}
  M.~Alishahiha and H.~Yavartanoo,
  ``Supergravity description of the large N noncommutative dipole field
  theories,''
  JHEP {\bf 0204}, 031 (2002)
  [arXiv:hep-th/0202131].


\bibitem{Alishahiha:2003ru}
  M.~Alishahiha and O.~J.~Ganor,
  ``Twisted backgrounds, pp-waves and nonlocal field theories,''
  JHEP {\bf 0303}, 006 (2003)
  [arXiv:hep-th/0301080].


\bibitem{Harmark:2000ff}
  T.~Harmark,
 ``Open branes in space-time non-commutative little string theory,''
  Nucl.\ Phys.\ B {\bf 593}, 76 (2001)
  [arXiv:hep-th/0007147].

\bibitem{Alishahiha:2000er}
  M.~Alishahiha,
   ``On type II NS5-branes in the presence of an RR field,''
    arXiv:hep-th/0002198.



\bibitem{Ho:2000fv}
  P.~M.~Ho and Y.~T.~Yeh,
   ``Noncommutative D-brane in non-constant NS-NS B field background,''
    Phys.\ Rev.\ Lett.\  {\bf 85}, 5523 (2000)
  [arXiv:hep-th/0005159].

\bibitem{Cornalba:2001sm}
  L.~Cornalba and R.~Schiappa,
   ``Nonassociative star product deformations for D-brane worldvolumes in
   curved backgrounds,''
    Commun.\ Math.\ Phys.\  {\bf 225}, 33 (2002)
  [arXiv:hep-th/0101219].

\bibitem{Ho:2001fi}
  P.~M.~Ho and S.~P.~Miao,
   ``Noncommutative differential calculus for D-brane in non-constant B  field
   background,''
    Phys.\ Rev.\ D {\bf 64}, 126002 (2001)
  [arXiv:hep-th/0105191].

\bibitem{Herbst:2001ai}
  M.~Herbst, A.~Kling and M.~Kreuzer,
   ``Star products from open strings in curved backgrounds,''
    JHEP {\bf 0109}, 014 (2001)
  [arXiv:hep-th/0106159].

\bibitem{Cerchiai:2003yu}
  B.~L.~Cerchiai,
   ``The Seiberg-Witten map for a time-dependent background,''
    JHEP {\bf 0306}, 056 (2003)
  [arXiv:hep-th/0304030].

\bibitem{Calmet:2003jv}
  X.~Calmet and M.~Wohlgenannt,
   ``Effective field theories on non-commutative space-time,''
    Phys.\ Rev.\ D {\bf 68}, 025016 (2003)
  [arXiv:hep-ph/0305027].

\bibitem{Bertolami:2003nm}
  O.~Bertolami and L.~Guisado,
   ``Noncommutative field theory and violation of translation invariance,''
    JHEP {\bf 0312}, 013 (2003)
  [arXiv:hep-th/0306176].

\bibitem{Robbins:2003ry}
  D.~Robbins and S.~Sethi,
   ``The UV/IR interplay in theories with space-time varying
   non-commutativity,''
    JHEP {\bf 0307}, 034 (2003)
  [arXiv:hep-th/0306193].

\bibitem{Behr:2003qc}
  W.~Behr and A.~Sykora,
   ``Construction of gauge theories on curved noncommutative spacetime,''
    Nucl.\ Phys.\ B {\bf 698}, 473 (2004)
  [arXiv:hep-th/0309145].

\bibitem{Das:2003kw}
  A.~Das and J.~Frenkel,
   ``Kontsevich product and gauge invariance,''
    Phys.\ Rev.\ D {\bf 69}, 065017 (2004)
  [arXiv:hep-th/0311243].

\bibitem{Gayral:2005ih}
  V.~Gayral, J.~M.~Gracia-Bondia and F.~Ruiz Ruiz,
   ``Position-dependent noncommutative products: Classical construction and
   field theory,''
    Nucl.\ Phys.\ B {\bf 727}, 513 (2005)
  [arXiv:hep-th/0504022].



\bibitem{Dolan:2002px}
  L.~Dolan and C.~R.~Nappi,
   ``Noncommutativity in a time-dependent background,''
    Phys.\ Lett.\ B {\bf 551}, 369 (2003)
  [arXiv:hep-th/0210030].

\bibitem{Hashimoto:2002nr}
  A.~Hashimoto and S.~Sethi,
   ``Holography and string dynamics in time-dependent backgrounds,''
  Phys.\ Rev.\ Lett.\  {\bf 89}, 261601 (2002)
  [arXiv:hep-th/0208126].

\bibitem{Lowe:2003qy}
  D.~A.~Lowe, H.~Nastase and S.~Ramgoolam,
   ``Massive IIA string theory and matrix theory compactification,''
  Nucl.\ Phys.\ B {\bf 667}, 55 (2003)
  [arXiv:hep-th/0303173].

\bibitem{Hashimoto:2004pb}
  A.~Hashimoto and K.~Thomas,
   ``Dualities, twists, and gauge theories with non-constant
   non-commutativity,''
  JHEP {\bf 0501}, 033 (2005)
  [arXiv:hep-th/0410123].


\bibitem{Alishahiha:2002bk}
  M.~Alishahiha and S.~Parvizi,
 ``Branes in time-dependent backgrounds and AdS/CFT correspondence,''
  JHEP {\bf 0210}, 047 (2002)
  [arXiv:hep-th/0208187].

\bibitem{Cai:2002sv}
  R.~G.~Cai, J.~X.~Lu and N.~Ohta,
  ``NCOS and D-branes in time-dependent backgrounds,''
  Phys.\ Lett.\ B {\bf 551}, 178 (2003)
  [arXiv:hep-th/0210206].

\bibitem{Hashimoto:2005hy}
  A.~Hashimoto and K.~Thomas,
  ``Non-commutative gauge theory on D-branes in Melvin universes,''
  arXiv:hep-th/0511197.
  
\bibitem{Kontsevich:1997vb}
  M.~Kontsevich,
   ``Deformation quantization of Poisson manifolds, I,''
  Lett.\ Math.\ Phys.\  {\bf 66}, 157 (2003)
  [arXiv:q-alg/9709040].



\bibitem{Takayanagi:2001aj}
  T.~Takayanagi and T.~Uesugi,
  ``D-branes in Melvin background,''
  JHEP {\bf 0111}, 036 (2001)
  [arXiv:hep-th/0110200].

\bibitem{Takayanagi:2001gu}
  T.~Takayanagi and T.~Uesugi,
  ``Flux stabilization of D-branes in NSNS Melvin background,''
  Phys.\ Lett.\ B {\bf 528}, 156 (2002)
  [arXiv:hep-th/0112199].


\bibitem{Dudas:2001ux}
  E.~Dudas and J.~Mourad,
  ``D-branes in string theory Melvin backgrounds,''
  Nucl.\ Phys.\ B {\bf 622}, 46 (2002)
  [arXiv:hep-th/0110186].
   

\bibitem{Costa:1996zd}
  M.~S.~Costa and G.~Papadopoulos,
 ``Superstring dualities and p-brane bound states,''
  Nucl.\ Phys.\ B {\bf 510}, 217 (1998)
  [arXiv:hep-th/9612204].

 
\bibitem{Horowitz:1991cd}
  G.~T.~Horowitz and A.~Strominger,
   ``Black strings and P-branes,''
  Nucl.\ Phys.\ B {\bf 360}, 197 (1991).

\bibitem{Minwalla:1999xi}
  S.~Minwalla and N.~Seiberg,
 ``Comments on the IIA NS5-brane,''
  JHEP {\bf 9906}, 007 (1999)
  [arXiv:hep-th/9904142].





\bibitem{Susskind:1998dq}
  L.~Susskind and E.~Witten,
  ``The holographic bound in anti-de Sitter space,''
  arXiv:hep-th/9805114.
\bibitem{Peet:1998wn}
  A.~W.~Peet and J.~Polchinski,
  ``UV/IR relations in AdS dynamics,''
  Phys.\ Rev.\ D {\bf 59}, 065011 (1999)
  [arXiv:hep-th/9809022].

\bibitem{Maldacena:1998im}
  J.~M.~Maldacena,
  ``Wilson loops in large N field theories,''
  Phys.\ Rev.\ Lett.\  {\bf 80}, 4859 (1998)
  [arXiv:hep-th/9803002].
  
\bibitem{Rey:1998ik}
  S.~J.~Rey and J.~T.~Yee,
  ``Macroscopic strings as heavy quarks in large N gauge theory and  anti-de
  Sitter supergravity,''
  Eur.\ Phys.\ J.\ C {\bf 22}, 379 (2001)
  [arXiv:hep-th/9803001].





\end{thebibliography}
\end{document}